\documentclass[cits]{JINST}
\title{Full Geant4 and FLUKA Simulations of an e$-$LINAC for its Use in Particle Detectors Performance Tests}
\author{
 B. Alpat$^a$\thanks{Corresponding author}, 
 E. Pilicer$^b$, 
 L. Servoli$^a$, 
 M. Menichelli$^a$, 
 P. Tucceri$^a$, 
 M. Italiani$^c$, 
 E.  Buono$^c$, 
 and F. Di Capua$^d$\\
\llap{$^a$}Istituto Nazionale Fisica Nucleare - Sezione di Perugia,Via A. Pascoli, Perugia I-06132, Italy\\
\llap{$^b$}Department of Physics, Uluda\u{g} University, 16059, Bursa, Turkey\\
\llap{$^c$}Azienda Ospedaliera "S. Maria", Viale Tristano di Juannuccio, 1,05100, Terni, Italy\\
\llap{$^d$}MAPRAD Srl, Via C.Colombo 19/I, 06127, Perugia, Italy\\
E-mail: \email{behcet.alpat@pg.infn.it}
}
\abstract{
 In this work we present the results of full Geant4 and FLUKA simulations and comparison with dosimetry data of an electron LINAC of St. Maria Hospital located in Terni, Italy. 
 The facility is being used primarily for radiotherapy and the goal of the present study is the detailed investigation of electron beam parameters to evaluate the possibility to use the e$-$LINAC (during time slots when it is not used for radiotherapy) to test the performance of detector systems, in particular those designed to operate in space. 
 The critical beam parameters are electron energy, profile and flux available at the surface of device to be tested.
 The present work aims to extract these parameters from dosimetry calibration data available at the e$-$LINAC. 
 The electron energy ranges from 4 MeV to 20 MeV. The dose measurements have been performed by using an Advanced Markus Chamber which has a small sensitive volume.}
\begin{document}
\hyphenation{agra-wal}

\section{Introduction}
The e$-$LINAC in use at Azienda Ospedaliera di Terni (AOT) has an electron energy range from 4 MeV to  20 MeV which is of major interest in the field of performance testing of devices operating in space. 
The e$-$LINAC of AOT is intensively used for radiotherapy of the patients, however, with a special agreement and in cooperation with AOT we have investigated the possibility to use, during nights and the weekends, the e$-$LINAC to test particle detector devices and components.

For this work, we set up a full geometrical description of e$-$LINAC and its electron beam to reproduce the dosimetry calibration data widely available at AOT. 
Using the Geant4 and FLUKA packages \cite{agostinelli03, allison06, battistoni07, fasso05} allows the full simulation of the  e$-$LINAC operations as well as to convert dosimetry data to parameters such as particle flux, energy spectrum and spot size at the surface of the devices positioned at various distances with respect to the e$-$LINAC head.

\section{Dose measurements}
\begin{figure}[th]
\centering
\includegraphics[width=.5\textwidth]{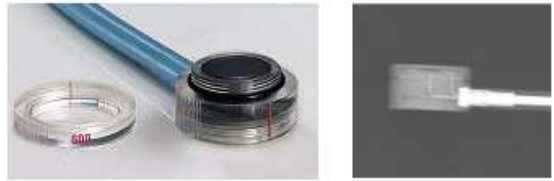}
\caption{Advanced Markus Chamber (AMC).}
\label{figure1}
\end{figure}

The Advanced Markus Chamber (AMC), which is of PTW type 34045 (its main parameters can be found in \cite{pearce06}), was used to perform dosimetry measurements in AOT. 
The Advanced Markus Chamber is a parallel plate ionization chamber and has a small sensitive volume, 0.02 cm$^3$, with a thin entrance window. The small sensitive volume of the chamber allows a dose distribution measurements in air and water, with good spatial resolution. The AMC used in the measurements can be seen in Figure~\ref{figure1}.

\begin{figure}[th]
\centering
\includegraphics[width=.5\textwidth]{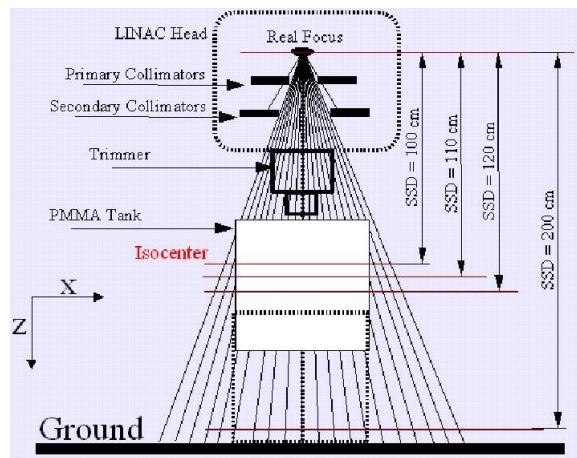}
\caption{A sketch of experimental setup used to take calibration data with Advanced Markus Chamber (AMC).  }
\label{figure2}
\end{figure}

In AOT, dose measurements have been performed at all available energies, ranging from 4 MeV to 20 MeV, with the trimmer (collimator) having a window surface of 3x3 cm$^2$. The AMC has been placed at distances here called Source to Surface Distance (SSD). The setup arranged for the dose measurements is shown in Figure~\ref{figure2}.

The AMC dose profiles have been evaluated at various distances (z--axis) and different energies in the transverse direction (x--axis) by measuring the FWHM of maximum electron beam intensity observed at the central axis position during an exposition time of one minute.

\section{Simulation}
The Geant4 and FLUKA simulation tools are developed primarily for use in High Energy Physics applications but are used in many areas such as space applications, medical applications and accelerator driven systems. We have developed simulation codes using the versions Geant4.9.3.p01 and Fluka2008.3c.

\begin{figure}[th]
\centering
\includegraphics[width=.5\textwidth]{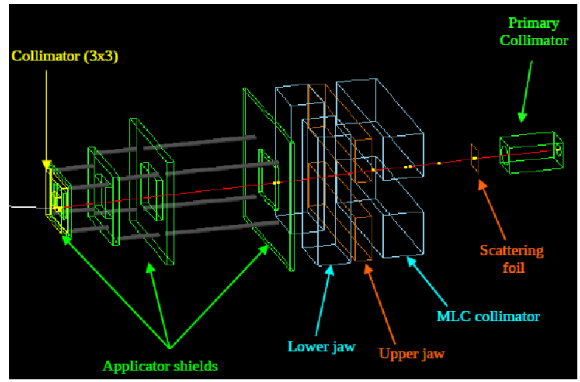}
\caption{Geometry of simulated acceleration head.}
\label{figure3}
\end{figure}

The geometry used in the simulation code consists of an accelerator head and small air ionization chambers placed along the scanning axis. 
The accelerator head shown in Figure~\ref{figure3} consists of internally embedded collimator leafs composed of a high density material and an applicator shield (trimmer) made of a high melting point Lipowitz material containing $50~\%$ bismuth, $26.7~\%$ lead, $13.3~\%$ tin and $10~\%$ cadmium by weight. 
Geant4 and FLUKA use different functions for geometrical definitions, but the description of the experimental setup and its operation is the same in both frameworks.

The primary electrons in both simulation codes were generated at a certain point and oriented along beam axis. Afterwards they hit a scattering foil followed by the collimator leafs inside the accelerator head and a trimmer. In order to obtain a preferred beam at the accelerator exit as foreseen in its manual the distance between collimator leafs has been kept flexible for tuning.

One of the key parameters to characterize the passage of a particle in a detector is its energy loss within the material. Since the particles of interest are electrons, we have used the Penelope physics, which is one of the recommended built-in physics converted from the PENELOPE code, for Geant4 and PRECISIO, which gives more accurate results with low particle-production and tracking thresholds, for FLUKA. Considering the relatively small thickness for the AMC sensitive region, where ionization takes place, a smaller step size has been applied for that part of the detector region. Detailed information about the underlying physics of particle interactions with matter has been implemented in our simulation frameworks can be found in Geant4 Physics Reference Manual  \cite{geant4} and FLUKA Online Manual  \cite{fluka}.

\subsection{Beam Profiles}
\begin{figure}[th]
\centering
\includegraphics[width=.85\textwidth]{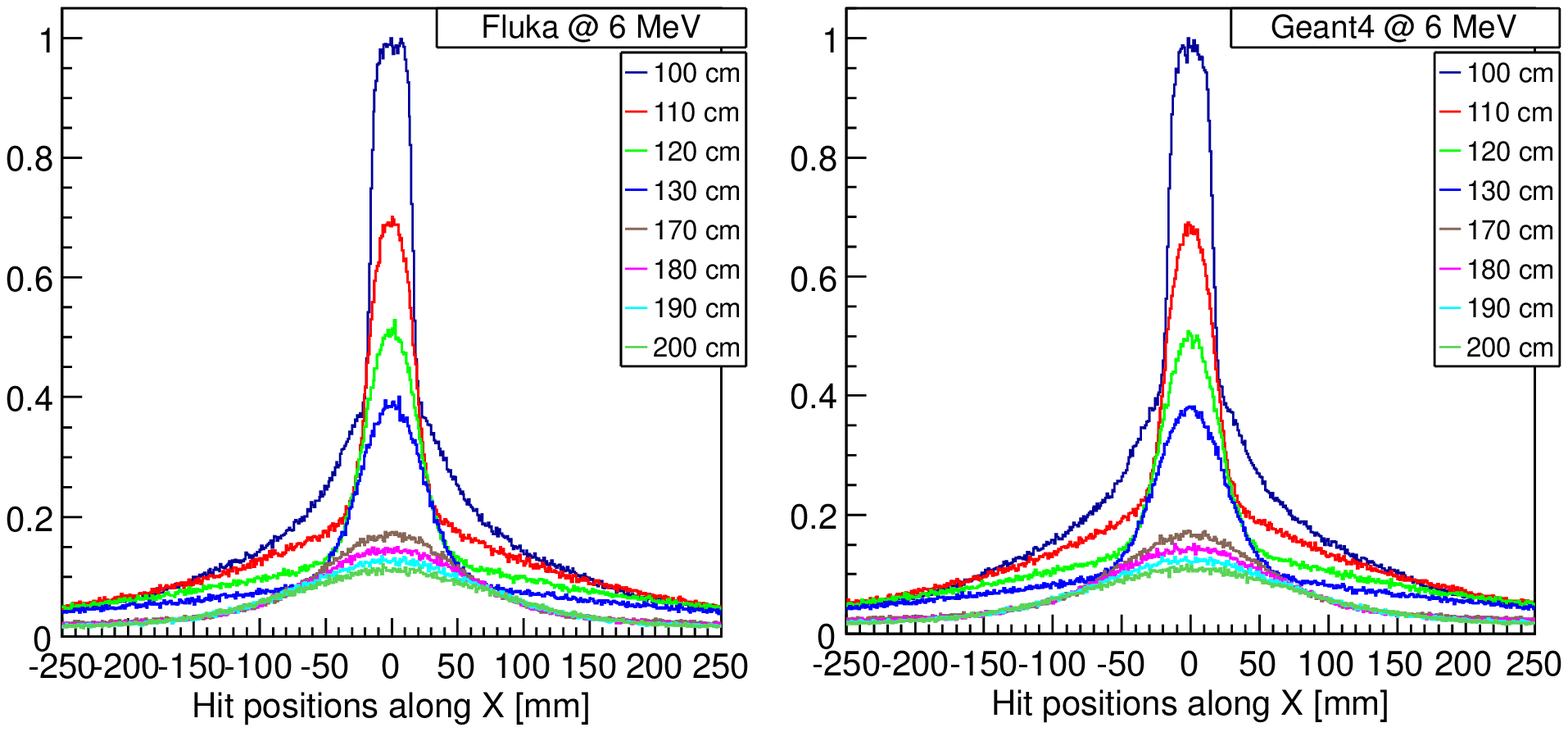}
\includegraphics[width=.85\textwidth]{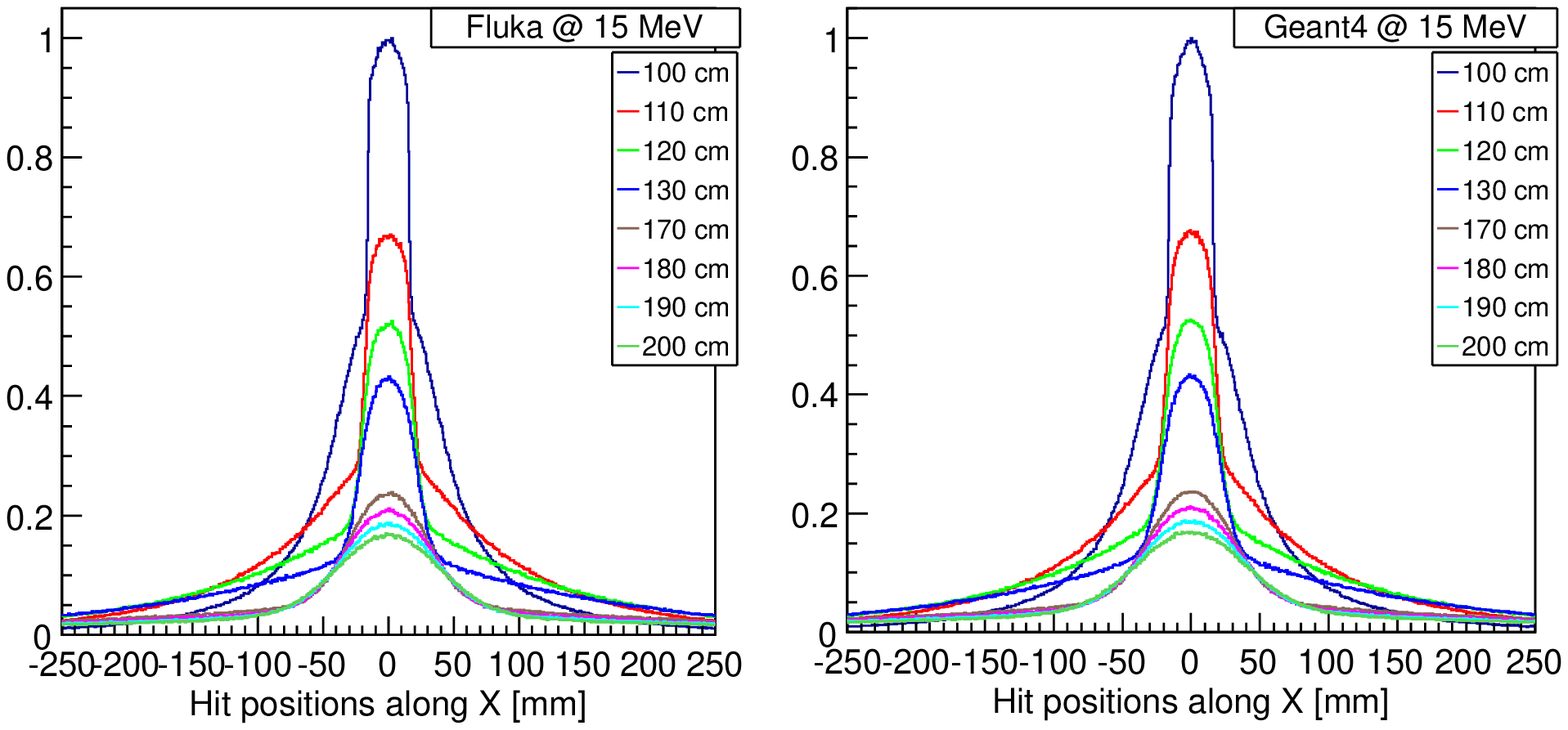}
\caption{
Change of beam profile widths at various exposition distances for all particles. 
}
\label{figure4}
\end{figure}

The beam profile FWHM widths measurements at a given SSD were performed by placing the AMC at positions perpendicular to the beam axis. 
As the SSD increases, due to small divergence of the beam and mostly because of multiple scattering, wider profiles were measured. 
Simulated variations of beam profile width spreads are plotted for two mid electron beam energies supported by the e$-$LINAC, at 6 MeV and 15 MeV, in Figure~\ref{figure4}. 
The shapes predicted by the two Monte Carlo simulations agree quite well.

A comparison of the FWHM measured by the AMC with the corresponding simulations from both Geant4 and FLUKA are presented in Table~I.

\subsection{Dose Profiles}
The Monte Carlo simulations directly predict the energy released into a specified volume, which is the AMC in this case. The relevant energy loss mechanism is the ionization in the chamber's sensitive volume. 

This deposited energy given in MeV in both simulations has been converted to the relative dose in order to compare it with the experimental data measured by the AMC. 
The relative dose values were then obtained at the central axis (x=0) by normalizing the dose values at a given SSD to the closest one at SSD=100 cm.
\begin{figure}[th]
\centering
\includegraphics[width=.85\textwidth]{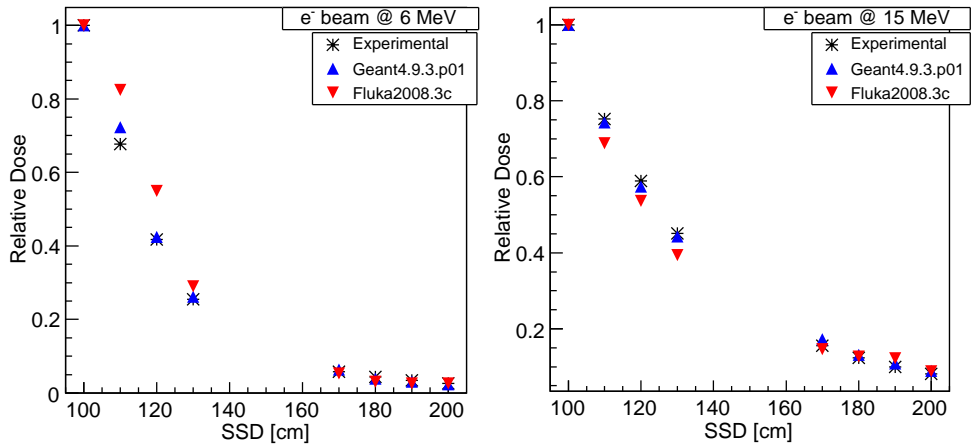}
\caption{
The relative dose normalized to the central axis (x=0)  at various exposition distances for 6 and 15 MeV. 
}
\label{figure5}
\end{figure}

The variations at two different beam energies, 6 MeV and 15 MeV, of the relative dose normalized to the one received at the central axis (x=0) are displayed in Figure~\ref{figure5}. 
Experimental data and Monte Carlo simulations are in good agreement over all exposition distances SSD. 
A smaller energy loss in the sensitive volume at larger distances is observed. 
In Table~II we compare the relative doses measured by the AMC and the estimated ones from the two Monte Carlo simulations.


\subsection{Kinetic Energy at the Surface}
We have shown that the beam and dose profiles accumulated in one minute at various conditions can be reproduced correctly. 
Thus, we can evaluate the kinetic beam energy, the other crucial beam parameter, at the detector surface for a given distance.
\begin{figure}[th]
\centering
\includegraphics[width=.85\textwidth]{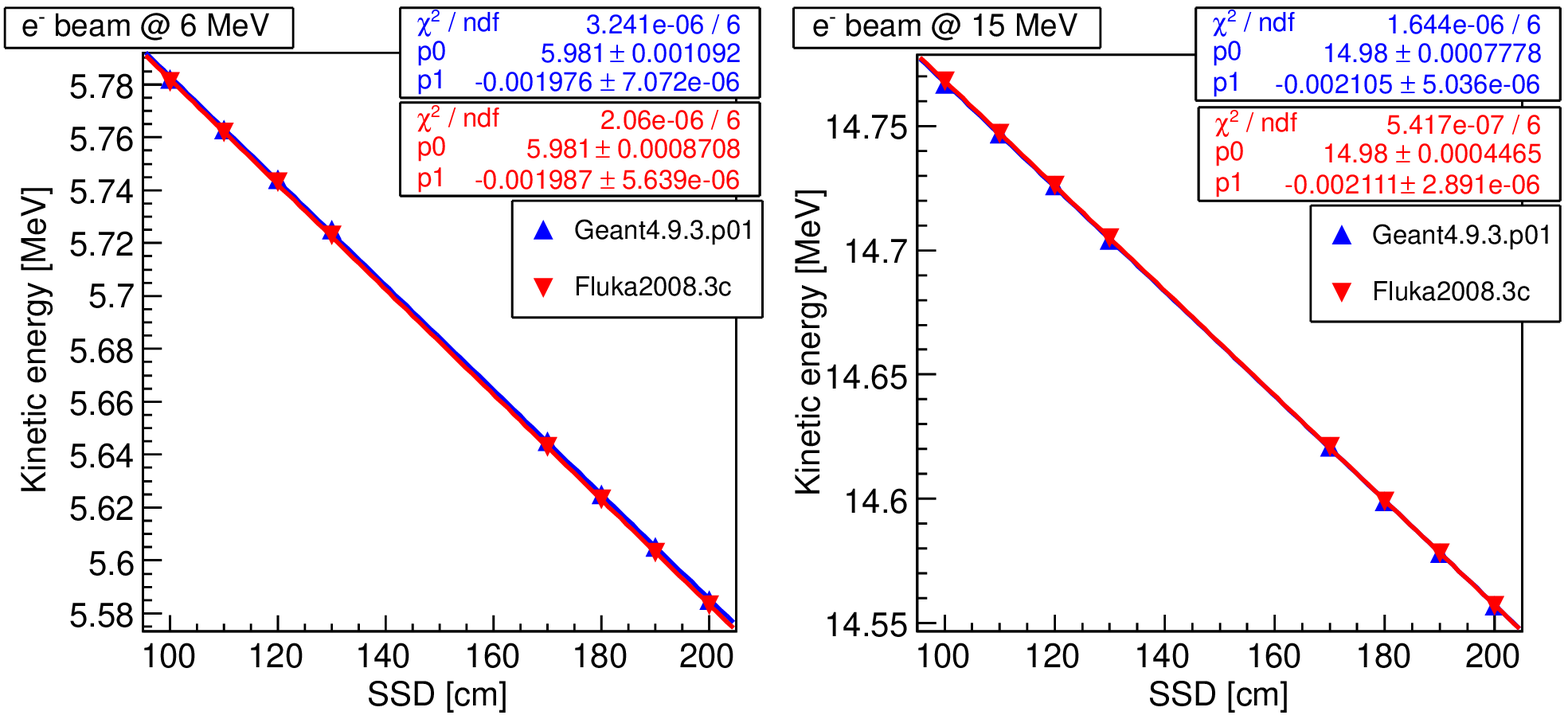}
\includegraphics[width=.85\textwidth]{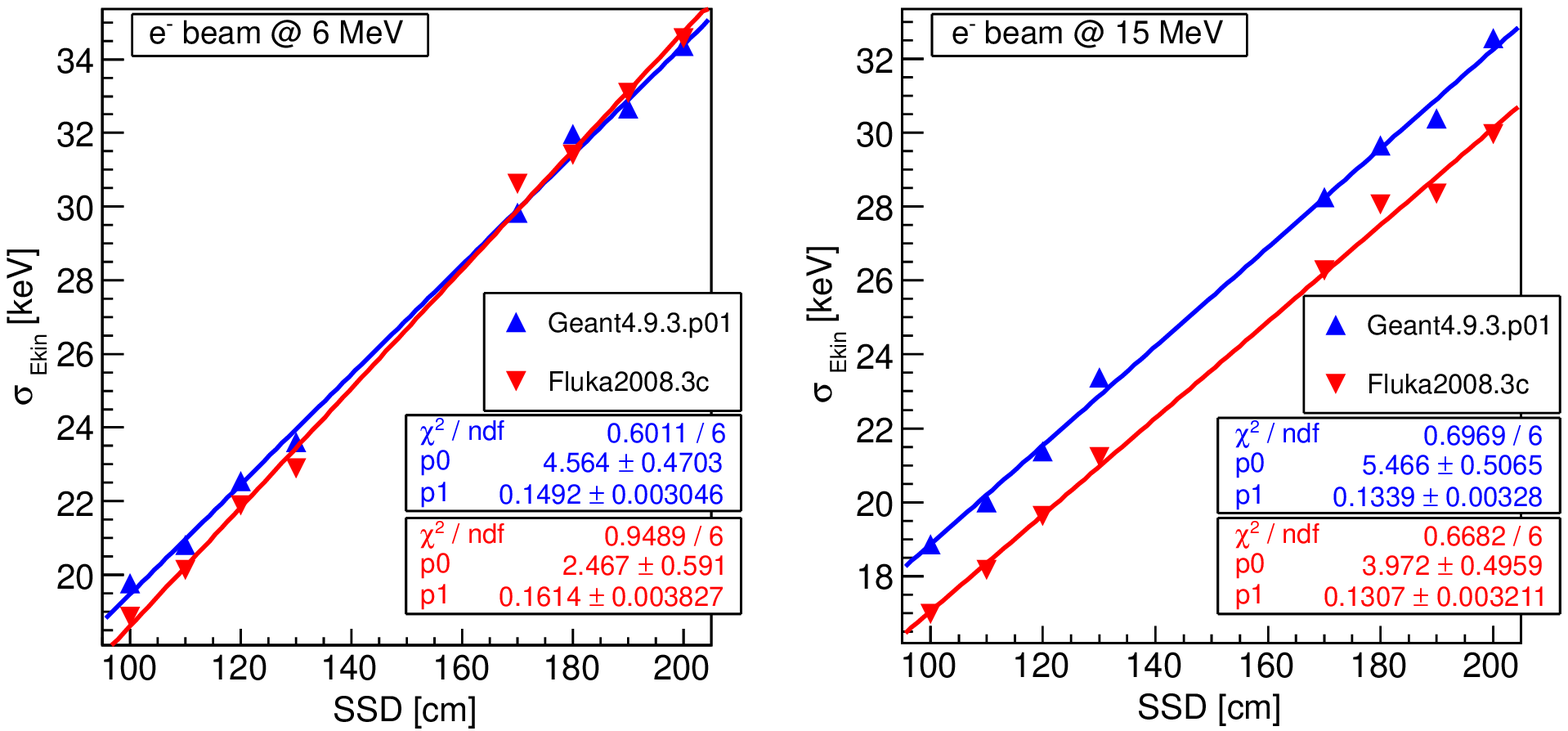}
\caption{
Kinetic beam energy and its fluctuations as a function of distance. 
}
\label{figure6}
\end{figure}
As the distance increase, the kinetic beam energy at the detector surface decreases while its fluctuation increases 
due to the energy spread. These two behaviors can be seen in Figure~\ref{figure6}. 

The knowledge of the kinetic beam energy and its spread is crucially important to tune the correct beam parameters 
(flux, energy and profile) for the detector under test.

\subsection{Conversion to the Flux}
The beam flux at a given distance can be evaluated by the following expression:
\begin{equation}
\mathrm{Flux} = \mathrm{CF} \times \mathrm{[~SurfaceHits / DetectorSurface / time~]}   
\end{equation}
where, the conversion factor (CF) is the factor to normalize dose obtained by simulation to the experimental dose values and time is given in one minute of radiation at a given distance from the AMC.
\begin{figure}[th]
\centering
\includegraphics[width=.85\textwidth]{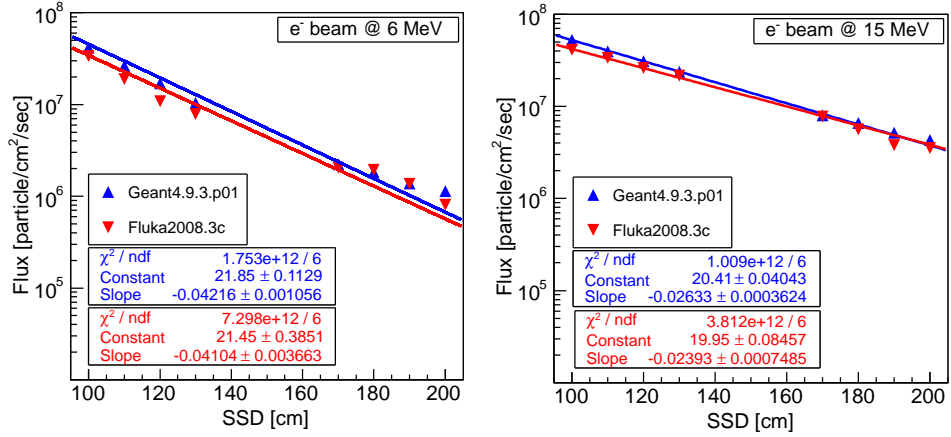}
\caption{
Electron beam fluxes as a function of SSD. 
}
\label{figure7}
\end{figure}
To study the performance of a particle detector at such an accelerator facility with high beam current may require reasonable extension in z direction to reduce the flux available on the detector surface to the desired level. 
From the fits performed in the simulations as illustrated in Figure~\ref{figure7}, one can extrapolate the expected lower particle flux at larger SSD.

\begin{figure}[th]
\centering
\includegraphics[width=.85\textwidth]{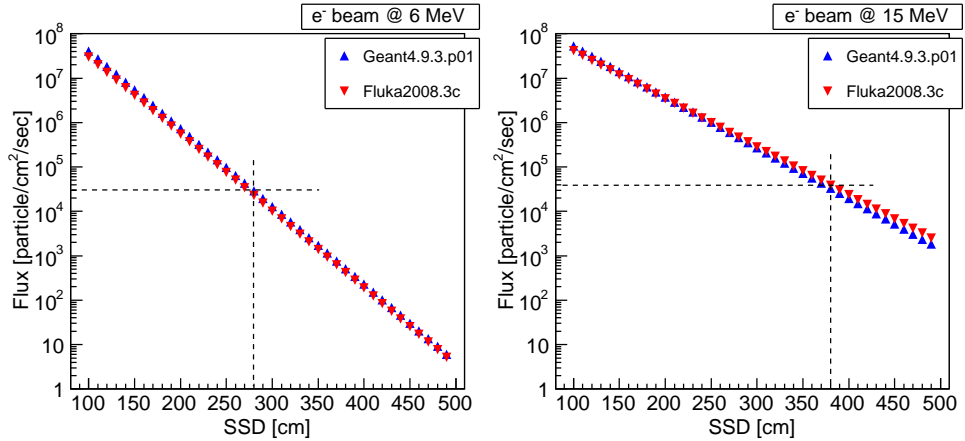}
\caption{
Predicted electron beam fluxes as a function of SSD. 
}
\label{figure8}
\end{figure}

The higher distances are necessary to reduce the beam flux for detector testing as can be seen in Figure~\ref{figure8}. 
For the electron beam energy of 6 MeV a proper distance to mount testing detector was found close to or larger than 250 cm while it is around 350 cm for the 15 MeV electron beam.

\section{Measurement with CMOS Pixel Detectors}
In addition to aforementioned studies based on conversions from available AMC data, an experimental measurement of the electron flux has also been performed using a CMOS pixel detector to count single electrons. 
Our group has investigated the charged particle detection capabilities for this kind of sensors in the past years \cite{meroli10, ozdemir11, sevoli08,  sevoli10}. 
The detection efficiency for such kind of detector is $99~\%$  for a minimum ionizing particle, so measuring the flux is a matter of counting the single electrons and normalizing it to the integration time and the detector surface.

\begin{figure}[th]
\centering
\includegraphics[width=.5\textwidth]{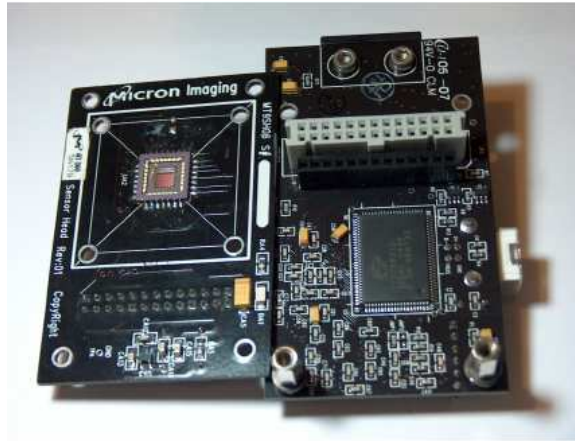}
\caption{Micron DAQ setup: the MT9V011 sensor with its DAQ board
(Demo2).}
\label{figure9}
\end{figure}

The CMOS sensor used in this case is a commercial VGA optical camera from Aptina Imaging, the MT9V011, as shown in  Figure~\ref{figure9}. 
It is a 640$\times$480 pixel matrix, with 5.6$\times5.6$ $\mu$m 
pixel size.

An electron passing through the sensor generates a signal shared only among few pixels (3$-$4 on average), even for small pixel size as in our case. The frame rate for the CMOS detector is 30 frames/second. The flux is computed as  number of detected electrons/sensor surface/integration time. 
All the pixels belong to the 3$\times$3 matrix centered on the pixel with the highest signal.
The clustering uses two thresholds algorithm; first threshold, to define the seed pixel,  requires the presence of the signal 10 times the single pixel noise, second threshold, to define the neighbor pixels, requires 2.5 times the single pixel noise applied to topologically connected pixels.   Hence it is a safe assumption to detect two different electrons if the related 3$\times$3 matrices are non$-$intersecting. This leads to a measuring capability of roughly $10~\%$ of the total number of pixels of the detector, which translates in our case to 3.5 $\times$ 10$^5$ electrons/cm$^2$/s.
\begin{figure}[th]
\centering
\includegraphics[width=.85\textwidth]{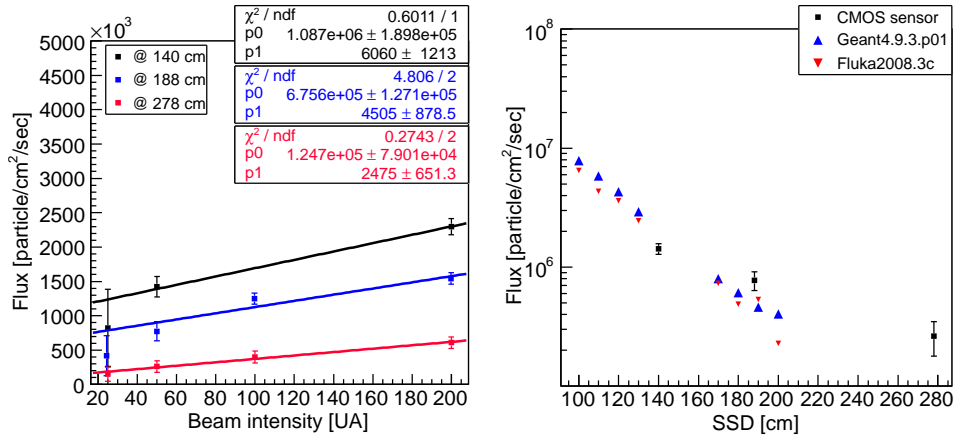}
\caption{
Predicted electron beam with kinetic energy of 10 MeV fluxes as a function of SSD. 
}
\label{figure10}
\end{figure}

The linearity of the measurement with the beam intensity has been calculated using the e$-$LINAC at electron energy of 10 MeV and 
at different distances. For beam intensities ranging from 25 U.A. up to 200 U.A., a linearity at the distances measured with 
CMOS sensor has been confirmed as seen in Figure~\ref{figure10} (left). 
The prediction of the Monte Carlo simulations compared to the measurements at available distances is presented in 
Figure~\ref{figure10} (right). Large error bar at 25 U.A. are caused by fluctuation of the beam. This value is a lower limit for the accelerator,
seldom used in clinical practice. The error bar at 140 U.A. is essentially dominated by statistical fluctuations, hence dominated by the Poissonian statistics where the absolute error is growing while the relative one is becoming smaller. However,  a good agreement within the errors has been achieved thereby validation the Monte Carlo setup.
\section{Conclusion}

The e$-$LINAC at (AOT) the Terni Hospital has a large experimental hall to perform detector testing with electrons of energies ranging from 4 MeV up to 20 MeV.
This work describes full Monte Carlo simulation of AOT's e$-$LINAC system and compares the resulting characteristics of the parameters like beam particle flux, kinetic energy and its fluctuations with experimental measurements. 
The dose measurements were done with an Advanced Markus Chamber (AMC) at different Source to Surface Distance (SSD) and with one minute radiation time at each distance. To compare with the experimental data, the simulations have been performed using two different simulation packages.
The simulations have shown that even with only AMC data one can extract relevant parameters such as beam energy, profile and flux, which are essential for detector performance studies. 
Moreover an additional measurement to validate the Monte Carlo results by CMOS pixels using measurement technique of single electron counting has also been done resulting in a good agreement between Monte Carlo predictions and experimental measurements.
During the data-simulation results checks, a systematic study was done to select the correct particle production thresholds (ppt) and  step sizes (ss) as well as physics options used in Geant4 simulations.  The PENELOPE package  was used (with ppt= 0.1 mm and ss=1 mm) since it has resulted to provide better agreement with data.

\newpage

\section{Appendix}
\begin{table}[ht]
\label{tab:table1}
\caption{Beam Profile Widths [cm] as a function of SSD [cm] for All Data Set Compared to MC.}
\centering
\vspace{1em}
\tiny\addtolength{\tabcolsep}{-5pt}
\begin{tabular}{|c|c|c|c|c|c|c|c|c|c|c|c|c|c|c|c|c|c|c|c|c|c|}
\hline
$E_\mathrm{kin}$ &\multicolumn{3}{c|}{4 MeV} &\multicolumn{3}{c|}{6 MeV} &\multicolumn{3}{c|}{8 MeV} &\multicolumn{3}{c|}{10 MeV} &\multicolumn{3}{c|}{12 MeV} &\multicolumn{3}{c|}{15 MeV} &\multicolumn{3}{c|}{18 MeV}\\ 
\hline
SSD &Data &Geant4 &Fluka &Data &Geant4 &Fluka &Data &Geant4 &Fluka &Data &Geant4 &Fluka &Data &Geant4 &Fluka &Data &Geant4 &Fluka &Data &Geant4 &Fluka\\
\hline
$100$ &$3.42$ &$3.21$ &$3.26$ &$3.39$ &$3.55$ &$3.47$ &$3.38$ &$3.60$ &$3.55$ &$3.38$ &$3.66$ &$3.67$ &$3.38$ &$3.66$ &$3.67$ &$3.38$ &$3.66$ &$3.67$ &$3.38$ &$3.66$ &$3.67$\\
\hline
$110$ &$4.07$ &$3.74$ &$3.63$ &$3.90$ &$4.23$ &$4.05$ &$3.80$ &$4.15$ &$4.08$ &$3.77$ &$4.25$ &$4.15$ &$3.38$ &$3.66$ &$3.67$ &$3.38$ &$3.66$ &$3.67$ &$3.38$ &$3.66$ &$3.67$\\
\hline
$120$ &$5.41$ &$5.50$ &$4.90$ &$4.72$ &$4.95$ &$4.49$ &$4.41$ &$4.78$ &$4.36$ &$4.24$ &$4.80$ &$4.72$ &$3.38$ &$3.66$ &$3.67$ &$3.38$ &$3.66$ &$3.67$ &$3.38$ &$3.66$ &$3.67$\\
\hline
$130$ &$7.08$ &$6.76$ &$6.34$ &$5.87$ &$6.51$ &$6.04$ &$5.28$ &$5.61$ &$5.27$ &$4.95$ &$5.35$ &$5.09$ &$3.38$ &$3.66$ &$3.67$ &$3.38$ &$3.66$ &$3.67$ &$3.38$ &$3.66$ &$3.67$\\
\hline
$170$ &$15.34$ &$13.70$ &$14.88$ &$12.16$ &$11.94$ &$11.20$ &$10.20$ &$10.36$ &$9.42$ &$9.13$ &$9.57$ &$9.33$ &$3.38$ &$3.66$ &$3.67$ &$3.38$ &$3.66$ &$3.67$ &$3.38$ &$3.66$ &$3.67$\\
\hline
$180$ &$17.09$ &$18.57$ &$17.87$ &$13.39$ &$12.78$ &$12.30$ &$11.22$ &$11.82$ &$11.43$ &$9.94$ &$10.60$ &$9.80$ &$3.38$ &$3.66$ &$3.67$ &$3.38$ &$3.66$ &$3.67$ &$3.38$ &$3.66$ &$3.67$\\
\hline
$190$ &$18.68$ &$19.07$ &$19.00$ &$14.59$ &$14.09$ &$14.20$ &$12.18$ &$13.64$ &$12.75$ &$10.75$ &$11.88$ &$11.68$ &$3.38$ &$3.66$ &$3.67$ &$3.38$ &$3.66$ &$3.67$ &$3.38$ &$3.66$ &$3.67$\\
\hline
$200$ &$20.94$ &$19.68$ &$19.32$ &$16.16$ &$16.58$ &$15.13$ &$13.29$ &$15.60$ &$13.50$ &$11.66$ &$13.23$ &$12.90$ &$13.38$ &$13.66$ &$13.67$ &$13.38$ &$13.66$ &$13.67$ &$13.38$ &$13.66$ &$13.67$\\
\hline
\end{tabular}
\end{table}

\begin{table}[ht]
\label{tab:table2}
\caption{Relative Dose Values as a function of SSD [cm] for All Data Set Compared to MC}
\centering
\vspace{1em}
\tiny\addtolength{\tabcolsep}{-5pt}
\begin{tabular}{|c|c|c|c|c|c|c|c|c|c|c|c|c|c|c|c|c|c|c|c|c|c|}
\hline
$E_\mathrm{kin}$ &\multicolumn{3}{c|}{4 MeV} &\multicolumn{3}{c|}{6 MeV} &\multicolumn{3}{c|}{8 MeV} &\multicolumn{3}{c|}{10 MeV} &\multicolumn{3}{c|}{12 MeV} &\multicolumn{3}{c|}{15 MeV} &\multicolumn{3}{c|}{18 MeV}\\ 
\hline
SSD &Data &Geant4 &Fluka &Data &Geant4 &Fluka &Data &Geant4 &Fluka &Data &Geant4 &Fluka &Data &Geant4 &Fluka &Data &Geant4 &Fluka &Data &Geant4 &Fluka\\
\hline
$100$ &$1.000$ &$1.000$ &$1.000$ &$1.000$ &$1.000$ &$1.000$ &$1.000$ &$1.000$ &$1.000$ &$1.000$ &$1.000$ &$1.000$ &$1.000$ &$1.000$ &$1.000$ &$1.000$ &$1.000$ &$1.000$ &$1.000$ &$1.000$ &$1.000$\\
\hline
$110$ &$0.598$ &$0.574$ &$0.643$ &$0.677$ &$0.723$ &$0.824$ &$0.720$ &$0.706$ &$0.686$ &$0.742$ &$0.729$ &$0.838$ &$0.752$ &$0.748$ &$0.786$ &$0.752$ &$0.743$ &$0.688$ &$0.748$ &$0.762$ &$0.771$\\
\hline
$120$ &$0.322$ &$0.292$ &$0.482$ &$0.418$ &$0.424$ &$0.549$ &$0.492$ &$0.484$ &$0.550$ &$0.541$ &$0.525$ &$0.542$ &$0.575$ &$0.561$ &$0.596$ &$0.589$ &$0.574$ &$0.537$ &$0.599$ &$0.585$ &$0.583$\\
\hline
$130$ &$0.181$ &$0.160$ &$0.146$ &$0.255$ &$0.262$ &$0.290$ &$0.325$ &$0.328$ &$0.375$ &$0.379$ &$0.397$ &$0.408$ &$0.423$ &$0.427$ &$0.454$ &$0.451$ &$0.443$ &$0.394$ &$0.475$ &$0.468$ &$0.467$\\
\hline
$170$ &$0.035$ &$0.038$ &$0.031$ &$0.058$ &$0.064$ &$0.053$ &$0.081$ &$0.079$ &$0.103$ &$0.103$ &$0.113$ &$0.111$ &$0.130$ &$0.142$ &$0.128$ &$0.155$ &$0.172$ &$0.146$ &$0.186$ &$0.186$ &$0.218$\\
\hline
$180$ &$0.026$ &$0.024$ &$0.019$ &$0.044$ &$0.039$ &$0.031$ &$0.061$ &$0.060$ &$0.057$ &$0.079$ &$0.089$ &$0.097$ &$0.102$ &$0.111$ &$0.108$ &$0.123$ &$0.131$ &$0.127$ &$0.150$ &$0.155$ &$0.168$\\
\hline
$190$ &$0.019$ &$0.017$ &$0.014$ &$0.034$ &$0.030$ &$0.026$ &$0.048$ &$0.051$ &$0.071$ &$0.062$ &$0.074$ &$0.057$ &$0.080$ &$0.092$ &$0.080$ &$0.099$ &$0.109$ &$0.122$ &$0.122$ &$0.127$ &$0.146$\\
\hline
$200$ &$0.016$ &$0.017$ &$0.010$ &$0.026$ &$0.024$ &$0.026$ &$0.037$ &$0.037$ &$0.007$ &$0.050$ &$0.051$ &$0.081$ &$0.064$ &$0.070$ &$0.071$ &$0.081$ &$0.089$ &$0.088$ &$0.101$ &$0.106$ &$0.115$\\
\hline
\end{tabular}
\end{table}

\bibliographystyle{JHEPdoi}
\bibliography{reference3}
\end{document}